\documentclass[aps,pra,preprint,showpacs,floatfix]{revtex4}

\usepackage{amsmath}
\usepackage{amssymb}
\usepackage{graphicx}
\usepackage{bm}
\usepackage{mathrsfs}

\newcommand{\ppi}{\bm{\mathnormal{\Pi}}}

\newcommand{\msl}{\mathscr{L}}
\newcommand{\msf}{\mathscr{F}}
\newcommand{\msg}{\mathscr{G}}

\begin{document}

\title{Propagation of Partially Coherent Photons \\ in an Ultra-Intense
  Radiation Gas}   

\author{Mattias Marklund}
\email{marklund@elmagn.chalmers.se}
\affiliation{Department of Electromagnetics,
  Chalmers University of Technology, SE--412 96 G\"oteborg, Sweden}
\affiliation{Institut f\"ur Theoretische Physik IV, Fakult\"at f\"ur
  Physik und Astronomie,  Ruhr-Universit\"at Bochum, D--44780 Bochum,
  Germany} 

\date{\today}

\begin{abstract}
  The scattering of photons off photons at the one-loop level 
  is investigated. We give a short review of the weak field limit, as
  given by the first order term in the series expansion of the
  Heisenberg--Euler Lagrangian. The dispersion relation for a photon
  in a radiation gas is presented. Based on this, a wave kinetic
  equation and a set of fluid equations is formulated. These equations
  describe the interaction between a partially coherent electromagnetic
  pulse and an intense radiation gas. The implications of the
  results are discussed.       
\end{abstract}

\maketitle

\section{Introduction}

Photon--photon scattering is a non-classical effect arising in
quantum electrodynamics (QED) due to virtual electron--positron
pairs in vacuum. Effectively, the interaction 
between photons and these virtual pairs will result in what is 
known as elastic photon--photon collisions. Formulated as an effective
field theory, using the Heisenberg--Euler Lagrangian
\cite{Heisenberg-Euler,Schwinger}, this results
in nonlinear corrections to Maxwell's vacuum equations, which to
lowest order in the fine structure constant are 
cubic in the electromagnetic (EM) field. These correction
takes the same form as in nonlinear optics, where the material
properties of, e.g., optical fibres, gives rise to cubic nonlinear
terms in Maxwell's equations, so called Kerr nonlinearities. 
Since the effective
self-interaction term in proportional to the fine structure
constant squared, this means that the field strengths need to
reach appreciable values until these effects becomes important
\cite{Greiner}.   
Higher order corrections can easily be incorporated, as will be done
here, by taking into 
account higher vertex and loop order diagrams. Possible detection
techniques \cite{bro01,bro02}, as well as physical implications, such
as the formation of light bullets
\cite{bro03,mar03,shu03,mar04,Shukla-etal}, of 
photon--photon scattering have attracted a rather constant
interest since first discussed (for a survey, both historical and
current, of the research in this area, see Refs.\ 
\cite{Greiner,Shukla-etal,Bialynicka,Ding} and references
therein). The concept of self-trapping 
of photons due to vacuum nonlinearities has also been discussed 
in the context of the one-dimensional
nonlinear Schr\"odinger equations \cite{Soljacic}. 
The non-trivial propagation of photons in strong background
electromagnetic fields, due to effects of nonlinear
electrodynamics, has been considered in a number of papers 
(e.g., Ref.~\cite{Bialynicka}). The main focus in
these papers has been on the photon splitting and birefringence
of vacuum, something which has attracted attention when it comes
to the extreme magnetic fields outside magnetars
\cite{Kouveliotou,Heyl}. It has even been suggested that the nontrivial
refractive index due to photon--photon scattering could induce a
lensing effect in the neighbourhood of a magnetar
\cite{Shaviv}. 
  
Here we will investigate the propagation of incoherent high frequency
photons on a radiation background, both of arbitrary intensities. It
will be shown that the nontrivial dispersion relation for a single
photon on a given background gives rise to novel collective nonlinear
effects. Furthermore, the general dispersion relation for a
perturbation in the system will be derived, and 
applications will be considered.

\section{Basic relations}

The effective field theory of one-loop photon--photon scattering in
constant background EM fields $\{\bm{E}, \bm{B}\}$ can be described by
the Lagrangian density $\msl = \msl_0 + \msl_{c}$ where $\msl_0 =
-\mathscr{F}$ is the classical free field Lagrangian, and 
\cite{Schwinger}  
\begin{eqnarray}
  \msl_{c} &=& -\frac{1}{8\pi^2}\int_0^{\mathrm{i}\infty}
  \frac{ds}{s^3}\mathrm{e}^{-m_e^2s} \Big[
  (es)^2ab\,\coth(eas)\,\cot(ebs)
\nonumber \\ &&  
  - \tfrac{1}{3}(es)^2(a^2 - b^2) - 1 \Big] ,
\label{eq:lagrangian}
\end{eqnarray}
where $m_e$ is the electron mass, $e = |e|$ is the electron charge, $a = 
[(\msf^2 + \msg^2)^{1/2} + \msf]^{1/2}$, $b = [(\msf^2 +
\msg^2)^{1/2} - \msf]^{1/2}$, $\msf \equiv F_{ab}F^{ab}/4 = 
(\bm{B}^2 - \bm{E}^2)/2$, $\msg \equiv
F_{ab}\widehat{F}^{ab}/4 = -\bm{E}\cdot\bm{B}$, 
$\widehat{F}^{ab} = \epsilon^{abcd}F_{cd}/2$. Thus, $\msf = (a^2 -
b^2)/2$ and $|\msg| = ab$. 

Following Ref.\ \cite{Marklund-Shukla-Eliasson2}, the above one-loop
Lagrangian yields the dispersion relation 
for a test photon in a background EM field of arbitrary field
strength. This can be formulated according to \cite{Dittrich-Gies2} 
\begin{eqnarray}
 &&\!\!\! 
  1 + \tfrac{1}{2}{\lambda}(\bm{E}^2 + \bm{B}^2) 
  + {\lambda}\Big[ -2(\hat{\bm{k}}\cdot{\ppi}) 
  \nonumber \\ && \!\!\!
  +
  \tfrac{1}{2}(\bm{E}^2 + \bm{B}^2) - 
  (\hat{\bm{k}}\cdot\bm{E})^2 - 
  (\hat{\bm{k}}\cdot\bm{B})^2\Big]n^2  = n^2,
\label{eq:disprel}
\end{eqnarray}
where $n = |\bm{k}|/\omega$ is the refractive index of the background,  
$\omega$ is the photon frequency, $\bm{k}$ is the photon wavevector,
and $\hat{\bm{k}} = \bm{k}/|\bm{k}|$. Here, we have used Abraham's
definition of the momentum density of the background EM
field \cite{Feigel}
\begin{equation}
  {\ppi} =  \frac{\bm{E}\bm{\times}\bm{B}}{n} .
\end{equation}
Furthermore, the ``effective
action charge'' ${\lambda}$ is given by  
\cite{Dittrich-Gies2} 
\begin{equation}
  {\lambda} = \frac{(\partial^2_{\msf} +
  \partial^2_{\msg})\msl}{-2\partial_{\msf}\msl +
  \msf(\partial^2_{\msf} + \partial^2_{\msg})\msl -
  2(\msf\partial^2_{\msf} + \msg\partial^2_{\msf\msg})\msl} .
\label{eq:Q}
\end{equation}

From Eqs.\ (\ref{eq:disprel}) and (\ref{eq:Q}) we see that as the
background EM fields vanish, so does the nonlinear effects, i.e., $n
\rightarrow 1$, as expected.

\section{Derivation of the governing equations}

Suppose that a plane wave pulse travels through an approximately
isotropic and homogeneous medium with refractive index $n$. The
relations $\bm{E}\cdot\bm{B} = 0$ and $|\bm{B}| = |\bm{E}| n$ then
holds. We note that if the background medium is the plane wave field
itself, the only physical solution to Eq.\ (\ref{eq:disprel}) is $n =
1$, i.e., a plane wave cannot self-interact. If the background
medium is a radiation gas, which we consider as an ensemble
$\{ \bm{E}_g,\bm{B}_g \}$ of plane
waves, the interaction contribution to the
dispersion relation (\ref{eq:disprel}) will be nonzero. We first note
that, since we are interested in the case of arbitrary intense fields,
we have to take the effect on each photon in both the pulse \emph{and}
the gas into account. The self-interaction within the gas will also be
nonzero, i.e., each photon in the gas will experience the refractive
index $n$ due to the gas and possible partially coherent plane wave
pulses. Thus, taking the ensemble average over the radiation gas
background in Eq.\ (\ref{eq:disprel}), we find the relation
\cite{Marklund-Shukla-Eliasson2} 
\begin{subequations}
\begin{eqnarray}
 n_g^2 = \frac{2 + {\lambda}_{g} \mathscr{E}}{1 -
  \tfrac{2}{3}{\lambda}_{g} \mathscr{E} + \sqrt{1 - 
  2{\lambda}_{g}\mathscr{E} +
  \tfrac{1}{9}({\lambda}_{g}\mathscr{E})^2}} 
\label{eq:refractive}
\end{eqnarray}
for the refractive index of a radiation gas, as experienced by a plane
wave pulse. 
The relation is valid for a \emph{weakly} anisotropic and
inhomogeneous radiation gas, since   
we have neglected the contribution from the averaged radiation gas
momentum density. Moreover, we have used 
$\langle(\hat{\bm{k}}_p\cdot\bm{E}_g)^2\rangle \simeq
\langle(\hat{\bm{k}}_p\cdot\bm{B}_g)^2\rangle/n^2 \simeq
\mathscr{E}/3$, where  $\mathscr{E} =
\langle|\bm{E}_g|^2\rangle$, and $\hat{\bm{k}}_p$ denotes the unit
wavevector of the pulse. In other words, the radiation gas is assumed
to be close to equilibrium. 
The relation (\ref{eq:refractive}) looks
deceptively simple, but one has to keep in mind that $\lambda_g =
\lambda_g(n_g,\mathscr{E})$  is determined through Eq.\
(\ref{eq:Q}). Thus, Eq.\ (\ref{eq:refractive}) constitutes an implicit
relation for the refractive index. 

In order to determine $\lambda_g$, we have to evaluate the derivatives
of the Lagrangian (\ref{eq:lagrangian}). For a radiation gas
background, where each photon experiences the radiation gas through
its refractive index, the invariants satisfy $a = [(n_g^2 - 1)
  \mathscr{E} ]^{1/2}$ and $b = 0$. Neglecting terms proportional to
$\alpha$ in the denominator of Eq.\ (\ref{eq:Q}), we obtain the
approximate expression 
\begin{equation}
  {\lambda}_g \simeq \frac{\alpha}{8\pi
  a^2}F(a/E_{\text{crit}}) . 
  \label{eq:Qgas}
\end{equation}
\label{eq:set1}
\end{subequations}
for the effective action charge. Here $F(a/E_{\text{crit}}) =
(4\pi/\alpha){a^2}(\partial^2_{\msf} +
\partial^2_{\msg})\msl_c|_{b = 0}$, and
$E_{\text{crit}} = m_e^2/e \sim 10^{16}\,\, \mathrm{V/cm}$ is
the Schwinger field \cite{Schwinger}.

\begin{figure}
\includegraphics[width=1\columnwidth]{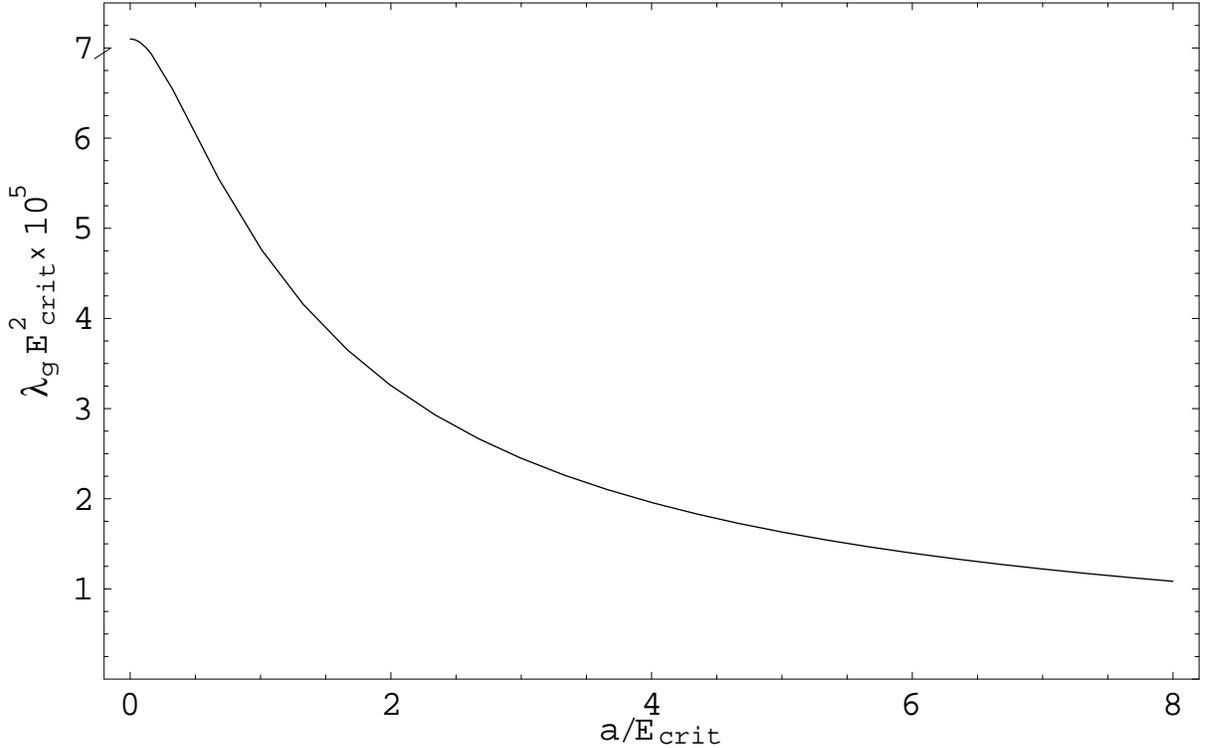}
\caption{The dimensionless function $(\alpha E_{\text{crit}}^2/8\pi
  a^2)F(a/E_{\text{crit}})\times 10^{5}$ plotted as a function of
  the dimensionless variable $a/E_{\text{crit}}$.} 
\label{fig:1}
\end{figure}

In Fig.\ \ref{fig:1} we have plotted the dimensionless number
$\lambda_g E_{\text{crit}}^2$ as a function of the dimensionless
variable $a/E_{\text{crit}}$. However, since $a$ is a function of $n$,
Eq.\ (\ref{eq:refractive}) still has to be solved for numerically in
order to obtain the refractive index as a function of the radiation
gas energy density $\mathscr{E}$. 

The case of a photon in random motion in the background field of a
plane wave pulse, represented by the slowly varying wave amplitude
$E_p$, can be treated in a similar way. Taking the ensemble 
average of Eq.\ (\ref{eq:disprel}) with respect to a random ensemble
of test photons yields 
\begin{subequations}
\begin{eqnarray}
 n_p^2 = \frac{2 + {\lambda}_{p} |E_p|^2}{1 -
  \tfrac{2}{3}{\lambda}_{p} |E_p|^2 + \sqrt{1 - 
  2{\lambda}_{p}|E_p|^2 +
  \tfrac{1}{9}({\lambda}_{p}|E_p|^2)^2}} .
\label{eq:refractive2}
\end{eqnarray}
Note that this can easily be generalised to incorporate the
self-interaction within the gas (which is non-zero in general). We
only need to make the replacement $|E_p|^2 \rightarrow
\mathscr{E}_{\text{tot}}$, where $\mathscr{E}_{\text{tot}} =
\mathscr{E} + |E_p|^2$. Equation (\ref{eq:refractive2}) has to be
supplemented by an expression for the effective action charge,
obtained from Eq.\ (\ref{eq:Q}) according to 
\begin{equation}
  {\lambda}_p \simeq \frac{\alpha}{8\pi
  a^2}F(a_p/E_{\text{crit}}) , 
  \label{eq:Qpulse}
\end{equation}
\label{eq:set2}
\end{subequations}
where $a_p = [(n_p^2 - 1) |E_p|^2 ]^{1/2}$. By the replacement
$|E_p|^2 \rightarrow \mathscr{E}_{\text{tot}}$ we also include the gas
self-interaction. 

In the weak field limit, $n_{p,g}$ is close to unity, and the
approximation $n_{p,g} \simeq 1$ may therefore be used in the
evaluation of ${\lambda}_{g,p}$. We then obtain  
${\lambda}_{g,p} \equiv \lambda =
(22/45)(\alpha/4\pi)E_{\text{crit}}^{-2}$. 
Keeping the first order terms in the energy densities of the pulse and
the gas, Eqs.\ (\ref{eq:refractive}) and (\ref{eq:refractive2}) yields
\begin{subequations}
\begin{equation}    
  n_g = 1 + \tfrac{2}{3}{\lambda} \mathscr{E} ,
\label{eq:weakrefractive}
\end{equation}
and 
\begin{equation}    
  n_p = 1 + \tfrac{2}{3}{\lambda} |E_p|^2 ,
\label{eq:weakrefractive2}
\end{equation}
\end{subequations} 
respectively. These results coincide with the ones obtained in Ref.\  
\cite{mar03}.

\subsection{Collective interactions}

The relation between the energy density $\mathscr{E}$ 
and the refractive index $n(\mathscr{E})$, as given by Eqs.\
(\ref{eq:set1}) or in the weak field limit by Eq.\
(\ref{eq:weakrefractive}), can be used to derive
a wave kinetic equation, determining the collective dynamics of a
partially coherent pulse of high frequency photons \cite{Mendonca}   
\begin{subequations}
\begin{equation}\label{eq:kinetic}
  \frac{\partial I}{\partial t}
  + \frac{1}{n_g(\mathscr{E})}\hat{\bm{k}}\cdot\bm{\nabla}I  
  + \frac{|\bm{k}|}{n_g(\mathscr{E})^2}\frac{d
  n_g(\mathscr{E})}{d\mathscr{E}}(\bm{\nabla}\mathscr{E})\cdot 
  \frac{\partial  
  I}{\partial\bm{k}} = 0 ,
\end{equation}
obtained using Hamilton's ray equations for the individual
photons. Here, the specific intensity $I(t, \bm{r}, \bm{k})$ is
normalised such that 
\begin{equation}
  \langle|E_p|^2\rangle = \int I\,d^3k 
\label{eq:normalised}
\end{equation}
\label{eq:kineticset}
\end{subequations}
is the energy density of the photons, and the angular brackets is the
ensemble average over the partially coherent photons in the pulse.  

Following Ref.\ \cite{mar03}, the dynamics of the low frequency
radiation gas, giving rise to the energy density $\mathscr{E}$, can be
formulated in terms of a coupled set of fluid equations. Assuming a
particle frame, and an 
equation of state $P = \mathscr{E}/3$, where $P$ is the radiation gas
pressure, the fluid hierarchy is closed at the second moment to yield 
\begin{subequations}
\begin{equation}\label{eq:energy}
  \frac{\partial\mathscr{E}}{\partial t} +
  \bm{\nabla}\cdot\left( \frac{\ppi}{n_p(\mathscr{E}_{\text{tot}})^2} 
  \right) =
  -\frac{\mathscr{E}}{n_p(\mathscr{E}_{\text{tot}})}\frac{d
  n_p(\mathscr{E}_{\text{tot}})}{d\mathscr{E}_{\text{tot}}}\frac{\partial 
  \mathscr{E}_{\text{tot}}}{\partial
  t} ,
\end{equation}
where $\mathscr{E}(\bm{r}, t)$ is weakly varying, and 
\begin{eqnarray}\label{eq:momentum}
  \frac{\partial{\ppi}}{\partial t} + \frac{1}{3}\bm{\nabla}\mathscr{E} =
  \frac{\mathscr{E}}{n_p(\mathscr{E}_{\text{tot}})}\frac{d
  n_p(\mathscr{E}_{\text{tot}})}{d\mathscr{E}_{\text{tot}}} \bm{\nabla}
  \mathscr{E}_{\text{tot}} , 
\end{eqnarray}
\label{eq:fluidset}
\end{subequations}
where $|\ppi|$ is small compared to $\mathscr{E}$, and
$\mathscr{E}_{\text{tot}} = \mathscr{E} + \langle|E_p|^2\rangle$. The
equations (\ref{eq:kineticset}) together with (\ref{eq:fluidset})
gives the collective interaction between partially coherent high
frequency photons and a radiation gas.

\section{Stability analysis}

We now turn our attention to the stability of perturbations of the
system (\ref{eq:kineticset}) and (\ref{eq:fluidset}). 
Starting with 
\begin{subequations}
\begin{equation}
I = I_0(\bm{k}) + I_1(t,\bm{r},\bm{k}), \,\, \mathscr{E} =
\mathscr{E}_0 + \mathscr{E}_1(t,\bm{r}),   
\end{equation}
and 
\begin{equation}
{\ppi} = {\ppi}_1(t,\bm{r}),
\end{equation}
\label{eq:perturbations}
\end{subequations}
where $I_0 \gg |I_1|$, $\mathscr{E}_0 \gg |\mathscr{E}_1|$
$\mathscr{E}_0 \gg |{\ppi}_1|$, we derive the general dispersion
relation for the linear perturbation. Inserting the form
(\ref{eq:perturbations}) into Eqs.\ (\ref{eq:kineticset}) and
(\ref{eq:fluidset}), we obtain
\begin{subequations}
\begin{eqnarray}
  && \!\!\!\!\!\!\!\!\!\!\!\!\!\!\!\!
  \frac{\partial I_1}{\partial t}
  + \frac{1}{n_{g0}}\hat{\bm{k}}\cdot\bm{\nabla}I_1  
  + \frac{|\bm{k}|}{n_{g0}^2}n_{g0}'(\bm{\nabla}\mathscr{E}_1)\cdot 
  \frac{\partial I_0}{\partial\bm{k}} = 0 , \\
  && \!\!\!\!\!\!\!\!\!\!\!\!\!\!\!\!
  \frac{\partial\mathscr{E}_1}{\partial t} +
  \frac{1}{n_{p0}^2}\bm{\nabla}\cdot\left({\ppi}_1 
  \right) =
  - \frac{\mathscr{E}_0}{n_{p0}}n_{p0}'\frac{\partial}{\partial
  t}(\mathscr{E}_1 + \langle|E_1|^2\rangle) ,
\end{eqnarray}
and 
\begin{equation}
  \frac{\partial{\ppi}_1}{\partial t} +
  \frac{1}{3}\bm{\nabla}\mathscr{E}_1 = 
  \frac{\mathscr{E}_0}{n_{p0}}n_{p0}' \bm{\nabla}
  (\mathscr{E}_1 + \langle|E_1|^2\rangle) ,
\end{equation}
\label{eq:pertset}
\end{subequations} 
where $n_{g0} = n_g(\mathscr{E}_0)$, $n_{p0} =
n_g(\mathscr{E}_{\text{tot},0})$, $n_{g0}' =
dn_g(\mathscr{E}_0)/d\mathscr{E}_0$, $n_{p0}' =
dn_g(\mathscr{E}_{\text{tot},0})/d\mathscr{E}_{\text{tot},0}$, and 
\begin{equation}
  \langle|E_{0,1}|^2\rangle = \int I_{0,1}\,d^3k .
\label{eq:integral}
\end{equation} 
Assuming a harmonic perturbation spectrum $\propto
\exp[i(\bm{K}\cdot\bm{r} - \Omega t)]$, Eqs.\ (\ref{eq:pertset}) and
(\ref{eq:integral}) yields the general dispersion relation
\begin{eqnarray}
  1 &=& -\varkappa n_{g0}'\frac{K^2 +
  n_{p0}^2\Omega^2}{\tfrac{1}{3}K^2(1 - 3\varkappa) -
  n_{p0}^2\Omega^2(1 + \varkappa)} 
  \nonumber \\ && \times   
  \int \frac{n_{g0}\Omega(\hat{\bm{k}}\cdot\bm{K}) + K^2 -
  2(\hat{\bm{k}}\cdot\bm{K})^2}{(n_{g0}\Omega -
  \hat{\bm{k}}\cdot\bm{K})^2}I_0 \,d^3k ,
\label{eq:pertdisprel}
\end{eqnarray}
for a perturbation of the system (\ref{eq:kineticset}) and
(\ref{eq:fluidset}). Here $\varkappa = \mathscr{E}_0n_{p0}'/n_{p0}$. 

In the case of a one-dimensional beam in the $z$-direction, $I_0 =
\langle|E_0|^2\rangle \delta^2(\bm{k}_{\perp})\delta(k_z - k_0)$, and
Eq.\ (\ref{eq:pertdisprel}) becomes
\begin{eqnarray}
  && \left[ \tfrac{1}{3}(K_{\perp}^2 + K_z^2)(1 - 3\varkappa) -
  n_{p0}^2\Omega^2(1 + \varkappa)\right](n_{g0}\Omega - K_z)^2
  \nonumber \\ && = 
  \varkappa n_{g0}' \langle|E_0|^2\rangle (K_{\perp}^2 + K_z^2 -
  n_{p0}^2\Omega^2)
  \nonumber \\ && \qquad\qquad\qquad\quad\times
  \left[ K_{\perp}^2 + K_z(n_{g0}\Omega - K_z)\right] .
\end{eqnarray}
We can see that the characteristics between longitudinal ($K_{\perp} =
0$) and transverse ($K_z = 0$) perturbations are very
different. The former is a stable perturbation, while the latter gives
an instability growth rate $\gamma = -i\Omega$ according to
\begin{eqnarray}
  &&\!\!\!\!\!\!\!\!\!\!\!\! \gamma^2 = \frac{K_{\perp}^2\left[
  \tfrac{1}{3}n_{g0}^2(1 - 
  3\varkappa) + \varkappa n_{g0}'\langle|E_0|^2\rangle
  \right]}{2n_{p0}^2n_{g0}^2(1 + 
  \varkappa)}
  \nonumber \\ &&\!\!\!\!\!\!\!\!\!\!\!\! \times
  \Bigg\{ \Bigg[ 1 + \frac{4n_{p0}^2n_{g0}^2n_{g0}'(1 +
  \varkappa)\varkappa \langle|E_0|^2\rangle}{\left[
  \tfrac{1}{3}n_{g0}^2(1 - 3\varkappa) + \varkappa
  n_{g0}'\langle|E_0|^2\rangle \right]^2} \Bigg]^{1/2} - 1\Bigg\} .
\end{eqnarray}

Furthermore, if the beam has a spread in $\bm{k}$-space, as
in the case of a Gaussian or Lorentzian distribution, there will
be damping due to the poles of the integrand in Eq.\
(\ref{eq:pertdisprel}) when the dimension of the physical system
exceeds one.

\section{Discussion and conclusions}

Currently, laser intensities can reach  
$10^{22}\,\, \mathrm{W/cm^{2}}$ and in the near future will go beyond
this value, possibly reaching $10^{24}\,\, \mathrm{W/cm^{2}}$
\cite{mou98}. The intensity limit of ordinary laser techniques
are well below the Schwinger critical
field strength \cite{mou98}, but laser-plasma
systems hold the promise of surpassing these limits, coming closer to
the Schwinger intensity $10^{29}\,\, \mathrm{W/cm^{2}}$
\cite{taj02,bul03}, at which the vacuum becomes fully nonlinear. 
When laser intensities reach the above field strengths, the plasma
particles will achieve highly relativistic quiver velocities. 
Thus, the relativistic ponderomotive
force will reach appreciable values, expelling the plasma particles
\cite{Yu-etal,shu04}. Due to this expulsion, plasma channels will form
and provide conditions for elastic photon--photon scattering
\cite{Shen-Yu,Shen-etal}. Thus, the next generation laser-plasma systems will
provide conditions at which Eqs.\ (\ref{eq:kineticset}) and
(\ref{eq:fluidset}) will be applicable.

In this paper, we have presented the equations describing the collective
interaction between a pulse of incoherent photons and a
radiation gas close to equilibrium, at arbitrary intensities. 
The derivation of these equations
was based on the one-loop Lagrangian of Schwinger, and the resulting
refractive index of the background electromagnetic fields. The
equations where linearised around a constant background, and the
general dispersion relation for the perturbation spectrum was
derived. It was shown that transverse perturbations are unstable,
and the corresponding instability growth rate was found. Applications
of the results to the next generation laser-plasma systems were
discussed. 

\acknowledgments
This work was supported by the European Commission (Brussels, Belgium)
through Contract No.\ HPRN-2001-0314.


\end{document}